\documentclass[final,a4paper,onecolumn,3p,12pt]{elsarticle}

\usepackage{graphicx,amssymb,mathrsfs,amsmath}
\usepackage{bm}
\usepackage{multirow}

\begin{document}
\title{Theoretical study of the $\Lambda$(1520) photoproduction off proton
target based on the new CLAS data}

\author[a,b,c]{Jun He}

\address[a]{Research Center for Hadron and CSR Physics,
Lanzhou University and Institute of Modern Physics of CAS, Lanzhou 730000, China}
\address[b]{Theoretical Physics Division, Institute of Modern Physics, Chinese Academy of Sciences,
Lanzhou 730000, China}
\address[c]{State Key Laboratory of Theoretical Physics, Institute of
Theoretical Physics, Chinese Academy of Sciences, Beijing  100190,China}

\begin{abstract}

Based on the high precision experimental data released by the
CLAS Collaboration recently, the interaction mechanism of
the photoproduction of $\Lambda$(1520) off a proton target is investigated
within a Regge-plus-resonance approach. With the decay amplitudes
predicted in the constituent quark model, the roles played by nucleon
resonances are studied. It is found that $N(2120)$ provides the most
important contribution among the nucleon resonances predicted in the
constituent quark model. The $t$ channel contribution with Regge
trajectories and the $\Lambda$ intermediate $u$ channel are
responsible to the behaviors of the differential cross section at
forward and backward angles, respectively.

\end{abstract}


\maketitle
\flushbottom

\section{Introduction}

Recently, the CLAS Collaboration at Jefferson National Accelerator
Facility released their exclusive photoproduction cross sections for
the $\Lambda$(1520), $\Sigma^0$(1385) and $\Lambda$(1405) for energies
from near threshold up to a center of mass energy $W$ of 2.85 GeV with
large range of the $K$ production angle~\cite{Moriya:2013hwg}
(labeled as CLAS13 in this work).  Since threshold for the
photoproduction of $\Lambda(1520)$ is about 2.01 GeV, the new
experimental data with high precision released by the CLAS
Collaboration provide an opportunity to study the nucleon resonances
above 2 GeV.

Among about two dozen nucleon resonances predicted by the constituent
quark model,a $D_{13}$ state $N(2120)$ with two star, which is labeled
as $N(2080)$ in the previous version of Particle Data Group (PDG)
\cite{PDG,Anisovich:2011fc}, should play the most important role in the photoproduction
of $\Lambda(1520)$ off proton target due to its large decay widths in
$\gamma p$ and $\Lambda(1520) N$ channels predicted in the constituent
quark model~\cite{Capstick:1998uh,Capstick:1992uc}.  $N(2120)$ has
attracted much attentions due to its importance found in many
channels, such as $\gamma p\to K^*\Lambda$~\cite{Kim:2011rm}, $\phi$
photoproduction~\cite{Kiswandhi:2011cq} and $\eta'p/\eta p$
photoproduction~\cite{Zhang:1995uha,Nakayama:2005ts,Zhong:2011ht,He:2008uf}.
A new bump structure was found at $W\simeq 2.1$ GeV in the
differential cross sections for the photoproduction of $\Lambda$(1520)
at forward angles measured by LEPS Collaboration at energies from near
threshold up to 2.4 GeV ~\cite{Kohri:2009xe}  (labeled as LEPS10), which
could be reproduced by including the resonance
$N(2120)$~\cite{Xie:2010yk,He:2012ud}.

In Refs.~\cite{Xie:2010yk,He:2012ud}, the photoproduction of
$\Lambda$(1520) has been investigated based on the LEPS10 data and the
differential cross sections are well reproduced.  However, only the
differential cross sections at forward angles are measured in the
LEPS10 experiment. At forward angles the $t$ channel contribution is
dominant while the $u$ channel contribution is negligible. Besides,
the contribution from $N$(2120) is expected to be important at all
kaon production angles.  Hence it is interesting to study the
interaction mechanism of the $\Lambda$(1520) photoproduction, especially
the $u$ channel and nucleon resonance intermediate $s$ channel, with the
CLAS13 data with large range of the kaon production angle. In a recent
work,  the $u$ channel contribution is found important at the backward
angles~\cite{Xie:2013mua}. However the differential cross sections at
forward angles are not so well reproduced.  In an analysis of the
CLAS13 data for $\Sigma$ (1385) photoproduction the Regge trajectory is found
essential to reproduce the behavior of the differential cross sections at
forward angles~\cite{He:2013ksa}. In this work we will study the
$\Lambda(1520)$ photoproduction off proton target in a
Regge-plus-resonance approach based on the new CLAS13 data to explore
the interaction  mechanism of the $\Lambda$ (1520) photoproduction.

This paper is organized as follows. After introduction, we will
present the effective Lagrangian and Regge trajectory used in this
work.  The experimental data will be fitted and the theoretical
results  of the differential and total cross sections compared with
experiment will be given in Sec.~\ref{Sec: Results}. Finally the
paper ends with a brief summary.

\section{Formula}

\begin{figure}[ht!]
	\begin{center}
\includegraphics[bb=130 590 420 720, scale=1.35,clip]{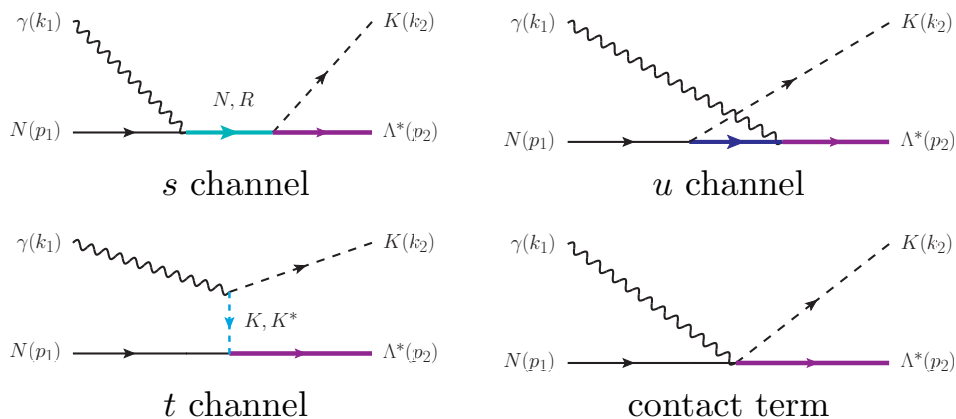}
\end{center}
\caption{(Color online) The diagrams for the $s$, $u$, $t$ channels
and the contact term.}
\label{pic:dia}
\end{figure}

The photoproduction of $\Lambda(1520)$  off proton target with
$K$ occurs through the following diagrams in Fig.~\ref{pic:dia}.

The Lagrangians use in the Born terms are given below~~\cite{Xie:2010yk,He:2012ud,Nam:2010au}
\begin{eqnarray}
\label{eq:GROUND}
{\cal L}_{\gamma KK}&=&
ieQ_K\left[(\partial^{\mu}K^{\dagger})K-(\partial^{\mu}K)K^{\dagger}
\right]A_{\mu},
\nonumber\\
{\cal L}_{\gamma
NN}&=&-e\bar{N}\left[Q_N\rlap{/}{A}-\frac{\kappa_N}{4M_N}\sigma^{\mu\nu}
F^{\mu\nu}\right]N,
\nonumber\\
\mathcal{L}_{KN\Lambda^*}&=&\frac{g_{KN\Lambda^*}}{M_{K}}
\bar{\Lambda}^{*\mu}\partial_{\mu}K\gamma_5N\,+{\rm H.c.}, \cr
{\cal L}_{\gamma
KN\Lambda^*}&=&-\frac{ieQ_Ng_{KN\Lambda^*}}{M_{K}}
\bar{\Lambda}^{*\mu} A_{\mu}K\gamma_5N+{\rm H.c.},
\end{eqnarray}
where $F^{\mu\nu}=\partial^\mu A\nu-\partial^\nu A_\mu$ with $A^\mu$,
$N$, $K$, $\Lambda^*$ are the fields for  the photon, nucleon, kaon and
$\Lambda^*(1520)$. $M_N$ and $M_K$ are the masses of the
nucleon and $K$ meson. $Q_h$ is
the charge of the hadron in the unit of $e=\sqrt{4\pi \alpha}$. The anomalous
magnetic momentum $\kappa_N=1.79$ for proton. With the decay width of
$\Lambda^*\to NK$ in Particle Data Group~(PDG)~\cite{PDG}, the
coupling constant can be obtained as $g_{KN\Lambda^*}=10.5$.

The $u$-channel diagram contains intermediate
hyperon $Y$ with spin-$/2$ as shown in Fig.~\ref{pic:dia}. The effective Lagrangians
for this diagram are
\begin{eqnarray}
\mathcal{L}_{KNY} &=& \frac{g_{KNY}^{}}{M_N +  M_Y} \overline{N} \gamma^\mu
\gamma_5^{} Y \partial_\mu K + \mbox{ H.c.},\nonumber\\
\mathcal{L}_{\Lambda^*Y\gamma} &=& -\frac{ief^u_1}{2M_Y} \overline{Y} \gamma_\nu
\gamma_5 F^{\mu\nu} \Lambda_\mu^*
+ \frac{ef^u_2}{(2M_Y)^2} \partial_\nu \overline{Y} \gamma_5 F^{\mu\nu}
\Lambda_\mu^* + \mbox{ H.c.},
\end{eqnarray}
where $F_{\mu\nu} = \partial_\mu A_\nu - \partial_\nu A_\mu$.
For the intermediate $\Lambda$ state, the coupling constants
$f^u_1 = -7.22$ and $f^u_2 = 9.48$ can be obtained from the ratio of
the helicity
amplitudes $A_{3/2}=-122$ and $A_{1/2}=13$ predicted in the constituent quark
model~\cite{Warns:1990xi}
and the PDG decay width $\Gamma_\gamma=133$ keV~\cite{PDG}. This
method has been
used in Ref.~\cite{Oh:2007jd} to determine the coupling constants for
$\Sigma(1385)\to \Lambda\gamma$.
The coupling constant $g_{KN\Lambda}$  can be determined by
flavor SU(3) symmetry, which gives value $g_{KN\Lambda}=
-13.24$~\cite{Oh:2007jd}. The $\Sigma$ exchange is negligible due to to the small
coupling constant determined from SU(3) symmetry~\cite{Oh:2007jd}.

It is well known that the amplitudes from nucleon intermediate $s$
channel, the $K$ exchange $t$ channel and the contact term are not gauge
invariant while for the sum of three amplitudes the
gauge invariance can be guaranteed. If the form factor which
reflects the hadron internal structure is included, the gauge
invariance will be violated. To restore the gauge invariance, a generalized contact
term is introduced as~\cite{Haberzettl:2006bn,Oh:2007jd}
\begin{eqnarray}
	M^{\mu\nu}_c&=&\frac{ieg_{KN\Lambda^*}}{m_K}\gamma_5\left[g^{\mu\nu}F_t+
		k_2^\mu
		(2k_2-k_1)^\nu\frac{(F_t-1)[1-h(1-F_s)]}{t-m_K^2}\right.\nonumber\\
		&+&\left.k_2^\mu
		(2p_1-k_1)^\nu\frac{(F_s-1)[1-h(1-F_t)]}{s-M_N^2}\right],
\end{eqnarray}
where the $h$ is free parameter and will be fitted by the experimental
data.

In this work, we adopt the form factor with form,
\begin{eqnarray}
F(q^2)&=&\left(\frac{n\Lambda^4}{n\Lambda^4+(q^2-M^2)^2}\right)^n,\label{Eq: FF}
\end{eqnarray}
where $M$ and $q$ are the mass and momentum of the off-shell
intermediate particle. In this work, except $n=2$ for $K^*$
exchange, we choose $n=1$ for other channels as Ref.~\cite{He:2013ksa}. The cut-off $\Lambda$ will be set as free parameter in this work.

It is well known that the behavior of the cross section for the
photoproduction at high photon energy is
described by Regge trajectory. We introduce the pseudoscalar and vector strange-meson Regge
trajectories following~\cite{Guidal:1997hy,Corthals:2006nz,Titov:2005kf}:
\begin{eqnarray}
\label{eq:RT}
\frac{1}{t-m^{2}_{K}}\to\mathcal{D}_{K}
&=&\left(\frac{s}{s_{scale}} \right)^{\alpha_{K}}
\frac{\pi\alpha'_{K}}{\Gamma(1+\alpha_{K})\sin(\pi\alpha_{K})},
\cr
\frac{1}{t-m^{2}_{K^{*}}}\to
\mathcal{D}_{K^{*}}&=&\left(\frac{s}{s_{scale}} \right)^{\alpha_{K^{*}}-1}
\frac{\pi\alpha'_{K^*}}{\Gamma(\alpha_{K^*})\sin(\pi\alpha_{K^*})},
\end{eqnarray}
where $\alpha'_{K,K^{*}}$ is the slope of the trajectory.  $\alpha_{K, K^{*}}$ 
is the linear trajectory, which is a function of $t$ as
$\alpha_{K}=0.70\,\mathrm{GeV}^{-2}(t-m^{2}_{K})$,
$\alpha_{K^{*}}=1+0.85\,\mathrm{GeV}^{-2}(t-m^{2}_{K^{*}})$.  The
scale constant $s_{scale}$ is set as 1 GeV$^2$. The coupling constant
for Regge trajectory
$\bar{g}^{Reg}_{KN\Lambda^*}=k^{R}g_{KN\Lambda^*}$ should be different
from the one for the real $K$ exchange
$g_{KN\Lambda^*}=10.5$~\cite{Xie:2010yk,He:2012ud,Nam:2010au}.  In
this work we set it as free parameter. We expect this difference
should not be very large. The $k^R$ for $K^*$ exchange is set to 1
considered that its contribution is small compared with $K$ exchange
contribution~\cite{He:2012ud}.

The Regge trajectories should work completely at high photon energies
and interpolate smoothly to a $K$ or $K^*$ exchange at low energies.
Toki $et\ al.$~\cite{Toki:2007ab} and Nam and Kao~\cite{Nam:2010au} introduced a weighting function to describe such
picture. Here we adopt the treatment as~\cite{He:2013ksa},
\begin{eqnarray}
\label{eq:R}
&&\frac{F_t}{t-m^{2}_{K}}\to \frac{F_t}{t-m^{2}_{K}} {\cal R}=\mathcal{D}_{K}R+\frac{F_t}{t-m^{2}_{K}}(1-{R}),\,\,\,\,
\end{eqnarray}
where $R={R}_{s}{R}_{t}$ with
\begin{eqnarray}
\label{eq:RSRT}
{R}_{s}=
\frac{1}{2}
\left[1+\tanh\left(\frac{s-s_{\mathrm{Reg}}}{s_{0}} \right)
	\right],\quad
	{R}_{t}=
1-\frac{1}{2}
\left[1+\tanh\left(\frac{|t|-t_{\mathrm{Reg}}}{t_{0}} \right) \right].
\end{eqnarray}
The free parameters $s_{Reg}$, $s_0$, $t_{Reg}$ and $t_0$ will be
determined by fitting experimental data.

The adoption of Regge trajectory will
violate the gauge invariance also. To restore the gauge invariance, we redefine the relevant amplitudes
as follow,
\begin{eqnarray}
\label{eq:WT1}
i\mathcal{M}_{t}+i\mathcal{M}_{s}+i\mathcal{M}_{c}
\to
i\mathcal{M}^{\mathrm{Regge}}_{t}+(i\mathcal{M}_{s}
+i\mathcal{M}_{c})\mathcal{R}.
\end{eqnarray}

The Lagrangians for the
resonances with arbitrary half-integer spin
are~\cite{He:2012ud,Oh:2007jd,Chang:1967zzc,Rushbrooke:1966zz,Behrends:1957},
\begin{eqnarray}
	\mathcal{L}_{\gamma N R(\frac{1}{2}^{\pm})} &=&\frac{e f_2}{2M_N}
	\bar{N} \Gamma^{(\mp)}\sigma_{\mu\nu}F^{\mu\nu} R \,+{\rm H.c.}, \\
\mathcal{L}_{\gamma N R(J^{\pm})} &=&\frac{-i^{n}f_1}{(2M_N)^{n}}
\bar{N}
~\gamma_\nu \partial_{\mu_2}\cdots\partial_{\mu_{n}}
F_{\mu_1\nu}\Gamma^{\pm(-1)^{n+1}}R^{\mu_1\mu_2\cdots\mu_{n}}\nonumber\\
&+&\frac{-i^{n+1}f_2}{(2M_N)^{n+1}} \partial_{\nu}\bar{N}
~ \partial_{\mu_2}\cdots\partial_{\mu_{n}}
F_{\mu_1\nu}\Gamma^{\pm(-1)^{n+1}}R^{\mu_1\mu_2\cdots\mu_{n}}+{\rm
H.c.},\label{Eq:Lg}
\end{eqnarray}
\begin{eqnarray}
\mathcal{L}_{RK\Lambda^*} &=& \frac{ig_2}{2m_{K}}
\partial_\mu K\bar{\Lambda}^*_{\mu} \Gamma^{(\pm)}R,+{\rm H.c.}, \\
\mathcal{L}_{RK\Lambda^*}
&=&\frac{-i^{n+1}g_1}{m_K^{n}} \bar{\Lambda}^*_{\mu_1}~\gamma_\nu\partial_\nu
\partial_{\mu_2}\cdots\partial_{\mu_{n}}
K\Gamma^{\pm(-1)^{n}}R^{\mu_1\mu_2\cdots\mu_{n}}\nonumber\\
&+&\frac{-i^{n}g_2}{m_K^{n+1}} \bar{\Lambda}^*_{\alpha}~\partial_{\alpha}\partial_{\mu_1}
\partial_{\mu_2}\cdots\partial_{\mu_{n}}
K\Gamma^{\pm(-1)^{n}}R^{\beta\mu_1\mu_2\cdots\mu_{n}}+{\rm H.c.},\label{Eq:Ls}
\end{eqnarray}
where $R_{\mu_1\cdots\mu_n}$ is the field for the
 resonance with spin $J=n+1/2$, and $\Gamma^{(\pm)}=(\gamma_5,1)$
for the different resonance parity. The coupling constants $f_1$,
$f_2$, $g_1$ and $g_2$ can be determined by the helicity amplitudes
$A_{1/2,3/2}$ and decay amplitudes $G(\ell_1,\ell_2)$ predicted in the
constituent quark model~\cite{Capstick:1998uh,Capstick:1992uc}. The interested reader is referred 
to Refs.~\cite{He:2012ud,He:2013ksa,Oh:2007jd}.

\section{Results}\label{Sec: Results}

With the formula given above, the CLAS13 and/or LEPS10  data for the
differential cross section of the $\Lambda(1520)$ photoproduction will be
fitted with the contributions from the nucleon and nucleon resonance
intermediate $s$ channels, the $K$ and $K^*$ exchange $t$ channels,
the $\Lambda$ intermediate $u$ channel and the contact term. The
fitting is done with the help of the MINUIT code in the cernlib.

\subsection{Fitting procedure}

In our model, the nucleon intermediate $s$ channel, $K$ and $K^*$
exchange $t$ channels with Regge trajectory, $\Lambda$ intermediate
$u$ channel and contact term are considered. The involved parameters
are listed in Table~\ref{Tab: parameters}. The best fitted values with
the uncertainties of
the parameters through fitting both CLAS13 and LEPS10 data are also
presented.
\begin{table}[h!]
\renewcommand\tabcolsep{0.25cm}
\renewcommand{\arraystretch}{1.3}
\begin{center}
\caption{The predetermined parameters and fitted parameters.
The mass, width and cut offs are in the unit of GeV.  The parameters for Regge
trajectory are in the unit of GeV$^2$.  \label{Tab: parameters}}
\begin{tabular}{lrlrlrlrlrlr}\hline
\multicolumn{8}{c}{ Predetermined parameters}\\\hline
 $g_{\gamma K K}$&$0.254$ & $g_{KN\Lambda^*}$&$10.5$
 &${g}_{K^*N\Lambda^*}$ & $20$ & $g_{KN\Lambda}$&$-13.24$\\
$f^u_1$ & -7.2 & $f^u_2$ & 9.5  &$\Gamma_R$ &330  &
  \\\hline\hline\multicolumn{8}{c}{ fitted parameters} \\\hline
  $\sqrt{s_{Reg}}$& $2.23\pm0.08$ & $s_0$ &$0.30\pm0.70$ 
  & $\sqrt{t_{Reg}}$ & $1.71\pm0.97$ & $t_0$ & $0.65\pm1.05$ 
  \\
 
  $\Lambda_{s,K}$ & $0.60\pm0.10$ & $\Lambda_u$ &$0.55\pm0.08$ 
  & $\Lambda_{K^*}$ &$0.79\pm0.64$ & $\Lambda_R$ & $0.70\pm0.25$ \\
 $k^R$ & $0.74\pm0.01$ & $h$ & $1.02\pm0.02$
	    \\\hline
\end{tabular}
\end{center}
\end{table}

In Table.~\ref{Tab: correlation}, the correlation coefficient of the
best fitted parameters is also presented. The $\Lambda_{s,K}$ and
$\Lambda_{K^*}$ have high correlation, which indicates the
contributions from $K$ and $K^*$ exchange play a similar role in the
reproduction of the experimental data.
\begin{table}[h!]
\renewcommand\tabcolsep{0.12cm}
\renewcommand{\arraystretch}{1.3}
\begin{center}
\caption{ Parameter correlation coefficients.  \label{Tab: correlation}}
\begin{tabular}{c|rrrrrrrrrr}\hline
&  $\sqrt{s_{Reg}}$& $s_0$& $\sqrt{t_{Reg}}$ & $t_0$
&  $\Lambda_{s,K}$& $h$  & $\Lambda_u$
  & $\Lambda_{K^*}$ & $\Lambda_R$ & $k^R$	    \\\hline
  $\sqrt{s_{Reg}}$  &$ 1.000$&$ 0.009$&$ 0.062$&$-0.052$&$-0.532$&$-0.371$&$-0.198$&$ 0.490$&$ 0.470$&$-0.003$\\\hline
  $s_0$&$ 0.009$    &$ 1.000$&$ 0.001$&$ 0.000$&$-0.009$&$-0.006$&$-0.005$&$ 0.008$&$ 0.013$&$ 0.011$\\\hline
 $\sqrt{t_{Reg}}$   &$ 0.062$&$ 0.001$&$ 1.000$&$ 0.507$&$-0.086$&$-0.521$&$ 0.042$&$ 0.088$&$-0.118$&$-0.039$\\\hline
  $t_0$             &$-0.052$&$ 0.000$&$ 0.507$&$ 1.000$&$ 0.057$&$ 0.180$&$-0.246$&$-0.093$&$ 0.212$&$-0.001$\\\hline
 $\Lambda_{s,K}$    &$-0.532$&$-0.009$&$-0.086$&$ 0.057$&$ 1.000$&$ 0.507$&$ 0.279$&$-0.983$&$-0.168$&$-0.005$\\\hline
  $h$               &$-0.371$&$-0.006$&$-0.521$&$ 0.180$&$ 0.507$&$ 1.000$&$-0.095$&$-0.526$&$-0.059$&$ 0.024$\\\hline
  $\Lambda_u$       &$-0.198$&$-0.005$&$ 0.042$&$-0.246$&$ 0.279$&$-0.095$&$ 1.000$&$-0.252$&$-0.442$&$-0.018$\\\hline
 $\Lambda_{K^*}$    &$ 0.490$&$ 0.008$&$ 0.088$&$-0.093$&$-0.983$&$-0.526$&$-0.252$&$ 1.000$&$ 0.106$&$ 0.004$\\\hline
  $\Lambda_R$       &$ 0.470$&$ 0.013$&$-0.118$&$ 0.212$&$-0.168$&$-0.059$&$-0.442$&$ 0.106$&$ 1.000$&$ 0.008$\\\hline
 $k^R$              &$-0.003$&$ 0.011$&$-0.039$&$-0.001$&$-0.005$&$ 0.024$&$-0.018$&$ 0.004$&$ 0.008$&$ 1.000$\\\hline
\end{tabular}
\end{center}
\end{table}

The nucleon resonances should be important at energies near
threshold of the  $\Lambda(1520)$ photoproduction. In the constituent
quark model a large amount of nucleon resonances in the
energy region considered in this work are predicted, which will make
the fitting very difficult. In this work we use the following criterion to select the
nucleon resonances which will be considered in the fitting,
\begin{eqnarray}
\lambda=(A_{1/2}^2+A_{3/2}^2)(G(\ell_1)^2+G(\ell_2)^2)\cdot10^{5}>0.01.
\end{eqnarray}
According to such criterion only six
resonances survive as listed in Table.~\ref{Tab: Resonances}.
Estimated from the values of $\lambda$, $N(2120)$ with
$\lambda=2.14$ should be the most important one among all nucleon
resonances considered in the current work.
\begin{table*}[h!]
\renewcommand\tabcolsep{0.405cm}
\renewcommand{\arraystretch}{1.3}
\caption{The nucleon resonances considered. The mass $m_R$, helicity
	amplitudes $A_{1/2,3/2}$ and partial wave decay amplitudes
	$G(\ell)$ are in the unit of MeV, $10^{-3}/\sqrt{\rm{GeV}}$ and
	$\sqrt{\rm{MeV}}$, respectively. The last column is for $\chi^2$ after
turning off the corresponding nucleon resonance and refitting. $\chi^2=3.05$ in
the full model. The amplitudes are from Refs.~\cite{Capstick:1998uh,Capstick:1992uc}.}
\begin{tabular}{ll|rrrr|rr}
 \hline State  &PDG & $A^p_{1/2}$ &  $A^p_{3/2}$ &
  $G(\ell_1)$ &  $G(\ell_2)$   & $\lambda$& $\chi^2$\\\hline
 $[N\textstyle{1\over 2}^-]_3(1945)$ & $N(1895)$ & 12  &  & 6.4 $^{+
 5.7}_{- 6.4}$ & & 0.59  & 3.04 \\
 $[N\textstyle{3\over 2}^-]_3(1960)$ &    $N(2120)$ & 36  & -43 & $-2.6 ^{+ 2.6}_{-
 2.8}$ & $-0.2 ^{+ 0.2}_{- 1.3}$ & 2.14 &3.54\\
 $[N\textstyle{5\over 2}^-]_2(2080)$ &$N(2060)$  & -3  & -14 & $-4.7 ^{+
 4.7}_{- 1.2}$ & $-0.3 ^{+ 0.3}_{- 0.8}$ &0.45
  &3.00  \\
  $[N\textstyle{5\over 2}^-]_3(2095)$ & & -2  & -6  & $-2.4 ^{+ 2.4}_{-
 2.0}$ & $-0.1 ^{+ 0.1}_{- 0.3}$ & 0.02  &3.05 \\
 $[N\textstyle{7\over 2}^-]_1(2090)$ &$N(2190)$ & -34  & 28  & $-0.5 ^{+ 0.4}_{- 0.6}$ & 0.0
 $^{+0.0}_{-0.0}$ & 0.05  &3.06\\
 $[N\textstyle{7\over 2}^+]_2(2390)$ & & -14  & -11  & 3.1 $^{+ 0.8}_{- 1.2}$ & 0.3 $^{+
 0.3}_{- 0.2}$ & 0.31   &3.14\\
 \hline
\end{tabular}
\label{Tab: Resonances}
\end{table*}

In the fitting procedure, the values of the nucleon resonance masses
suggested in PDG~\cite{PDG} are preferred. For the nucleon
resonances which is not listed in PDG, the prediction of the constituent
quark model will be
adopted. To avoid the proliferation of the free parameters, the
Breit-Wigner widths for all nucleon resonances are set to 330~MeV,
which is consistent to the values  suggested
in PDG for the nucleon resonances listed in Table.~\ref{Tab:
Resonances}~\cite{PDG}. Here, we would like to mention that with the
help of the helicity and decay amplitudes listed in Table.~\ref{Tab:
Resonances} the
contributions from nucleon resonances are only dependent on one free
parameter $\Lambda_R$, which is also presented in Table~\ref{Tab: parameters}.

With the parameters listed in Table~\ref{Tab: parameters}, the
experimental data about the differential cross section will be fitted.
In the fitting procedure, we only consider CLAS13 and LEPS10 data. The
reduced $\chi^2$ per degree of freedom are presented in
Table~\ref{Tab: chi2}. The corresponding values of fitted parameters
are listed in Table~\ref{Tab: parameters}.  Here only statistic
uncertainties are included in the fitting procedure. To compare with
Ref.~\cite{Xie:2013mua}, the results with systematic uncertainty
$\sigma_{sys}=11.6\%$ and 5.92\% for CLAS13 and LEPS10, are also
listed in Table~\ref{Tab: chi2}. The results with only CLAS13 data are
also presented for comparison.
\begin{table*}[h!]
\renewcommand\tabcolsep{0.37cm}
\renewcommand{\arraystretch}{1.3}
\caption{The reduced $\chi^2$ for the full model and the models after turning off $u$
channel, all nucleon resonances, the nucleon resonances except $N(2120)$ the
Regge trajectory or the $K^*$ exchange  and refitting. The second to forth columns are for the results
with both CLAS13 and LEPS10 data. The last three columns are for the
results with CLAS13 data only. The second (five) and third (sixth) columns
are for the results with and without systematic uncertainties
$\sigma_{sys}$. }
\begin{tabular}{l|ccc|ccc}
	\hline  &\multicolumn{3}{c|}{CLAS13\&LEPS10}
	&\multicolumn{3}{c}{CLAS13} \\
	\hline  & no $\sigma_{sys}$ & with $\sigma_{sys}$ &Ref.~\cite{Xie:2013mua} &
	no $\sigma_{sys}$ & with $\sigma_{sys}$ &Ref.~\cite{Xie:2013mua}\\\hline
	Full model   & 3.05 &[1.38]  &[2.5] &2.65& [1.07] & [2.5]\\
	no u channel  & 4.10&[2.08]&[9.9] & 4.08  & [2.13] &[5.6]\\
	no $N^*$       & 4.82&[2.12] &[3.0]& 4.79 & [2.22] &[3.0]\\
only $N(2120)$   & 3.27&[1.48] & & 2.85   & [1.08]  & \\
no Regge  & 8.39&[2.95] && 7.69 & [2.51]\\
no $K^*$  & 3.36&[1.51] && 2.89 & [1.18]\\
 \hline
\end{tabular}
\label{Tab: chi2}
\end{table*}

For the full model $\chi^2=3.05$ and decreases to 1.38 if the
systematic uncertainty are included, which is much smaller than
 $\chi^2=2.5$ in Ref.~\cite{Xie:2013mua}. If LEPS10 data is excluded,
 $\chi^2$ decreases to 2.65.  If turning off $\Lambda$
intermediate $u$ channel,  $\chi^2$
increases to 4.10, which suggests the importance of the $u$ channel
contribution. After excluding all nucleon
resonances, $\chi^2$ increases to 4.82. However, if  $N(2120)$ is
included, $\chi^2$ decreases to 3.27, which is close to 3.05 with full
model. The result shows that $N$(2120) is dominant in the
nucleon resonance contribution. The importance of $N(2120)$ can be
also  observed in Table~\ref{Tab:
Resonances}. After turning off $N(2120)$ the
corresponding $\chi^2$ is 3.54, which is much larger other nucleon
resonances as the value of $\lambda=2.14$ for $N(2120)$. We also check the
importance of the Regge trajectory. A value of $\chi^2$ about 8 is
found if the Regge trajectory is turned off. If the systematic
uncertainty are included, $\chi^2$ is $2.95$, which is close to the
$\chi^2=2.5$ in Ref.~\cite{Xie:2013mua}. Compared with $\chi^2=3.05$
for the full model, the Regge trajectory is essential to describe the
experimental differential cross section. To check the correlation of
the contributions from the $K$ exchange and $K^*$ exchange, the
$\chi^2$ after turning $K^*$ exchange are also presented. It is found
that the absence of $K^*$ contribution is compensated by the $K$
contribution as suggested by the high correlation between
$\Lambda_{s,K}$ and $\Lambda_{K^*}$ in Table~~\ref{Tab: correlation}.

\subsection{Differential cross section}

Compared with the reduced $\chi^2$, the explicit differential cross
section can provide more information about interaction
mechanism of $\Lambda(1520)$ photoproduction.  First, we present
our results for CLAS13 and CLAS10 data which are used in the fitting procedure.

\begin{figure}[h!]
\begin{center}
  \includegraphics[ bb= 170 240 700 618 ,scale=0.93,clip]{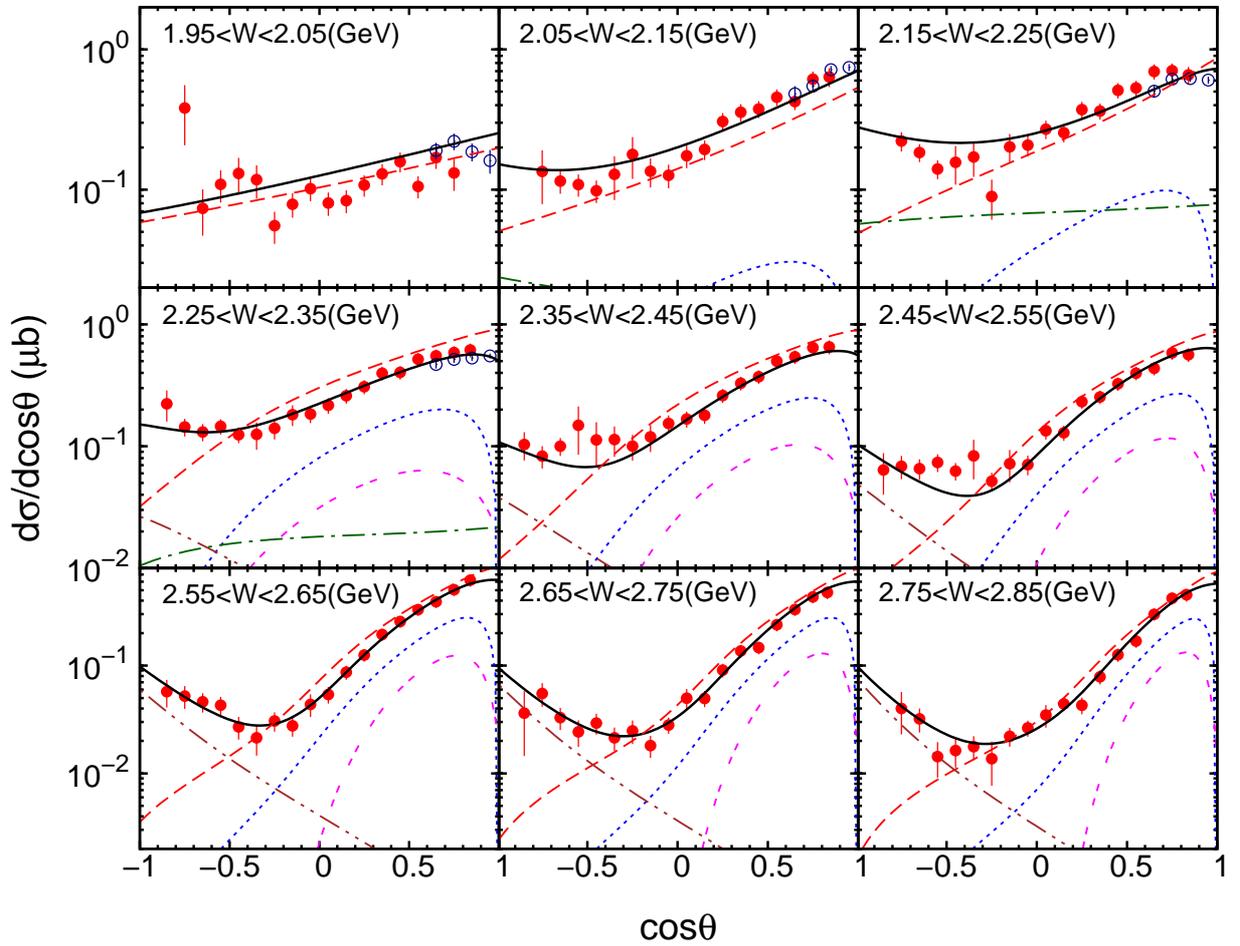}
  \caption{(Color online) The differential cross section
	  $d\sigma/d\cos\theta$ with variation
	  of $\cos\theta$. The full (black), dashed (red), dotted
  (blue), dash-dot-dotted( brown), dash-dashed (magenta) and
  dash-dotted (darkgreen) lines are for full model, contact term, $K$ exchange $t$
  channel, $u$ channel, $K^*$ exchange $t$ channel and $N(2120)$
  . The full circle (red) and open circle (blue) are for
  CLAS13 data \cite{Moriya:2013hwg} and LEPS10 data
  \cite{Kohri:2009xe}.}
  \label{Fig: CLAS13}
\end{center}
\end{figure}

In Fig.~\ref{Fig: CLAS13}, the differential cross sections compared with CLAS13 data are
figured. The LEPS10 data at same energies are also plotted in the
corresponding subfigure.
One can find that experimental data are well reproduced in our model at
all energies and angles. The dominant contributions
are from Born terms, among which the contact term plays the most important role.
At the low energy the $K$ exchange contribution is smaller and becomes
more important at high energies (please note here and hereafter that
the amplitude of the contact term, the $K$ exchange, or the $s$ channel is
not gauge invariance. Here, the result is obtained in the center of
mass frame).  The vector meson $K^*$ exchange
provides considerable contribution at high energies. Compared with Fig.3 in
Ref.~\cite{Xie:2013mua}, the Regge trajectory is important to
reproduce the behavior of the differential cross section at forward
angles especially at high energies. The $u$ channel is responsible to
the increase of the differential cross section with the decrease of
$\cos\theta$ at high energies.  However, the similar increase at low
energies is from $N(2120)$ contribution instead.

The results compared with LEPS10 data are figured in Fig.~\ref{Fig:
LEPS10dcs}. The CLAS13 data at same angles are also plotted.
\begin{figure}[h!]
\begin{center}
  \includegraphics[ bb=190 345 525 590,scale=1.36,clip]{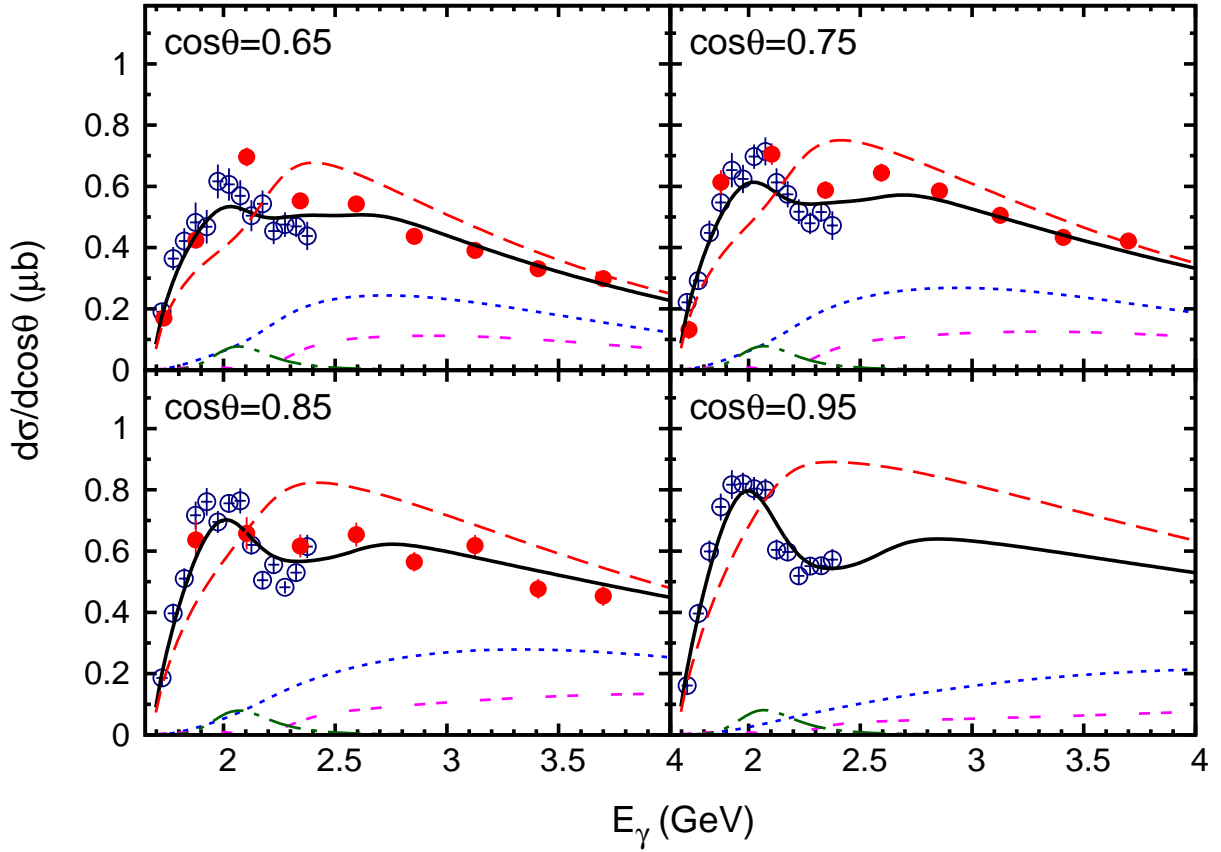}
\end{center}
  \caption{(Color online) The differential cross section
	  $d\sigma/d\cos\theta$ with the variation of the photon
	  energy $E_\gamma$ compared with LEPS10 data. The notation as Fig.~\ref{Fig: CLAS13}.}\label{Fig: LEPS10dcs}
\end{figure}
One can find that there exist discrepancies between LEPS10 data and CLAS13
data especially at $\cos\theta=0.65$. Our results are close to the
LEPS10 data and the bump structure is well reproduced with the
contribution from $N(2120)$. The contact term contribution is dominant
at energies from near threshold up to $E_\gamma=4$ GeV.  Generally, the
CLAS13 data is reproduced especially at high energies. Compared with
Fig.4 in Ref.~\cite{Xie:2013mua} a large difference can be found in
the prediction of the differential cross section at extreme forward
angle $\cos\theta=0.95$. A sharp increase was predicted at energies
larger than 2.5 GeV in Ref.~\cite{Xie:2013mua} while our result
suggests a smooth heave. The difference should be from adoption of
Regge trajectory in the current work which is not considered in
Ref.~\cite{Xie:2013mua}.  It can be checked in the future
experiment.

In the fitting, the polarization asymmetry measured in LEPS10
experiment are not included. As shown in the previous
work~\cite{Xie:2010yk,He:2012ud},the sign of the polarization
asymmetry should be reversed compared with the experiment. The results
in this work confirm the previous conclusion as shown in Fig.~\ref{Fig:
LEPS10POL}.
\begin{figure}[h!]
\begin{center}
  \includegraphics[ bb=0 268 390 510,scale=1.1,clip]{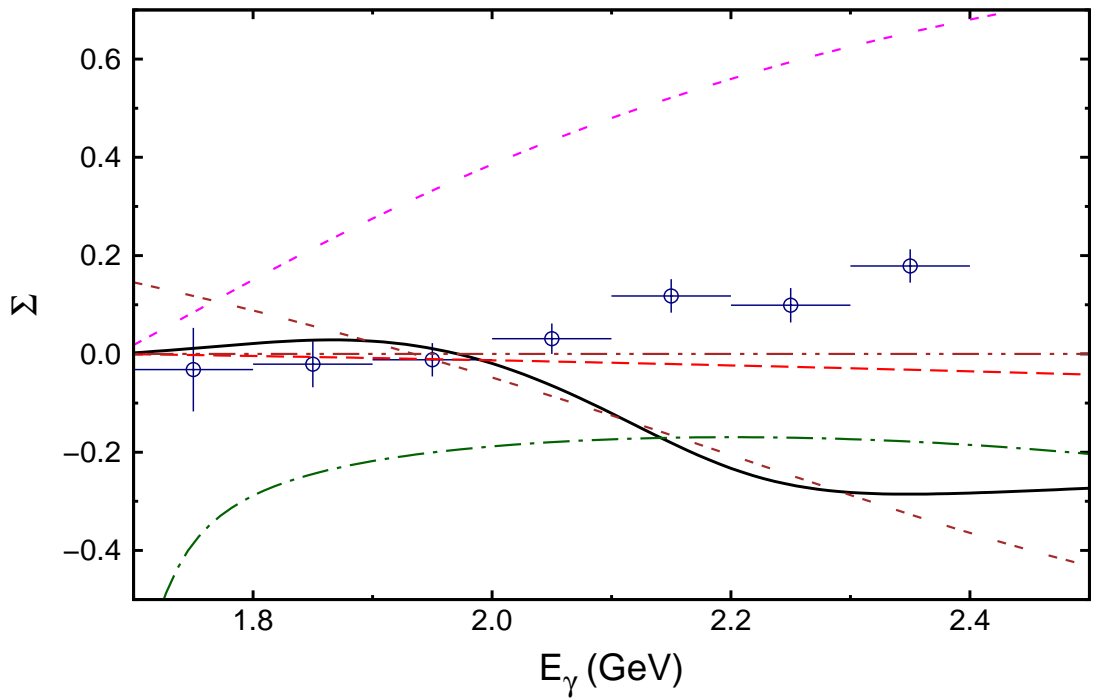}
\end{center}
  \caption{
	  (Color online) Polarization asymmetry with the
	  variation of the energy of photon $E_\gamma$
 compared with
LEPS10 data~\cite{Kohri:2009xe}. The notation as  Fig.~\ref{Fig: CLAS13}.}\label{Fig: LEPS10POL}
  \end{figure}

To show a picture around the resonance pole of $N(2120)$, the
$d\sigma/d\cos\theta$ against $\theta$ at 2.2~GeV is calculated and
compared with LEPS09
data in Ref.~\cite{Muramatsu:2009zp} which are not included in the fitting
procedure.
\begin{figure}[h!]
\begin{center}
\includegraphics[ bb=181  355 565 680,scale=1.2,clip]{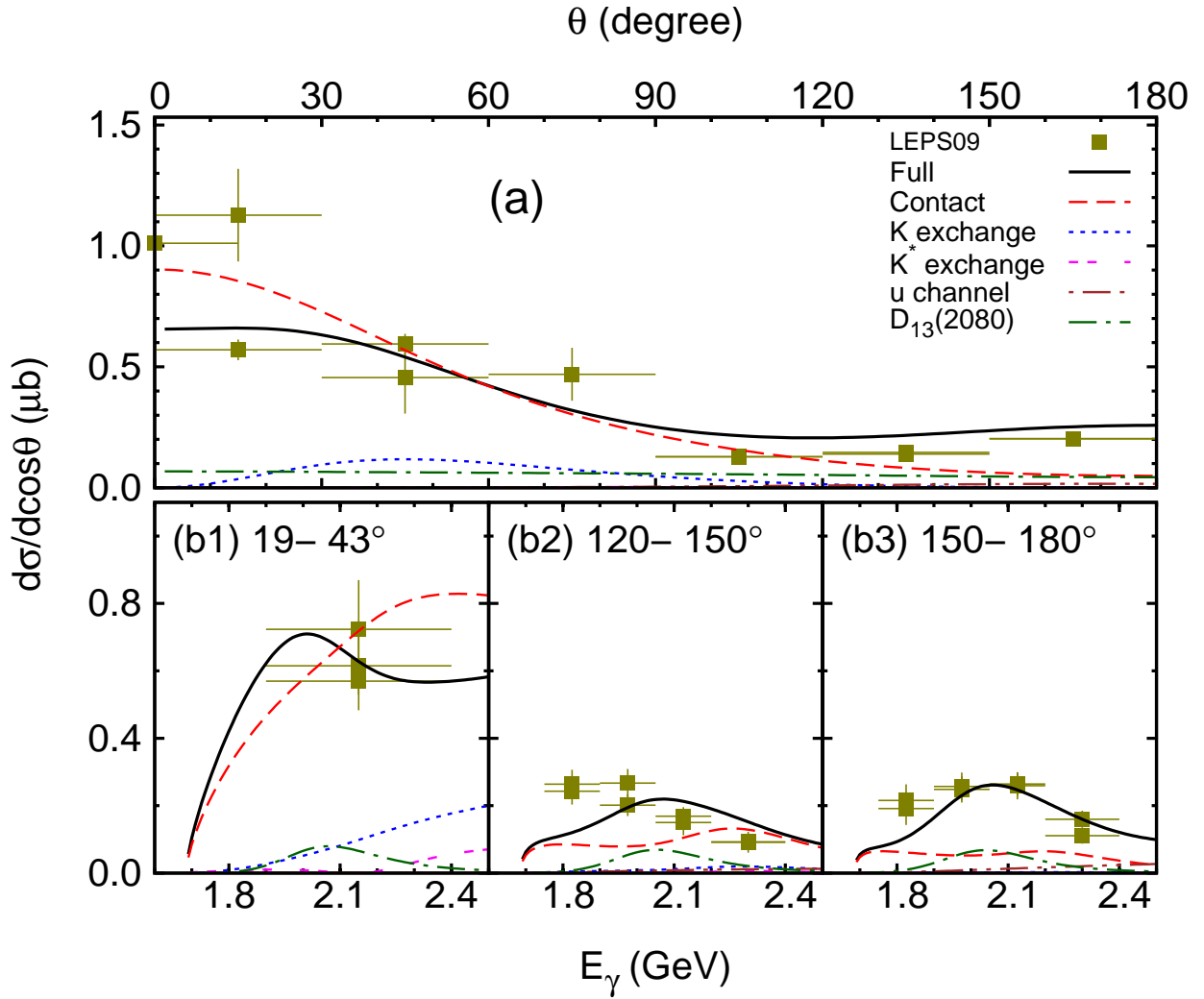}
\end{center}
\caption{(Color online) (a) The differential cross section
	$d\sigma/d\cos\theta$ with the variation of $\theta$ at
photon energy 2.2~GeV compared with the LEPS09 data at photon
energy 1.9-2.4GeV. (b1-b3) The differential cross section
$d\sigma/d\cos\theta$ with the variation of photon energy $E_{\gamma}$
.
the notation as in Fig.~\ref{Fig: CLAS13}. The experimental data are from
Ref.~\cite{Muramatsu:2009zp}.}\label{Fig: LEPS09dcs06}
\end{figure}
As shown in Fig.~\ref{Fig: LEPS09dcs06},
the experimental data are reproduced generally in our model.
The general shape for the differential cross section against
$\theta$ is mainly formed by the contact term contribution. The slow
increase at backward angles is from $N(2120)$ contribution. For
differential cross section with the variation of $E_\gamma$,
$N(2120)$ provides contribution comparable with the contact term
at backward angles and is responsible to the bump structure.

Recently CLAS
Collaboration reported their preliminary results about the
photoproduction of $\Lambda(2120)$ off both proton and neutron in
deuterium at the photon energy from $1.87$ to $5.5$~GeV
in the $eg3$ run~\cite{Zhao2011}. We do not consider
these preliminary $eg3$
data in the fitting procedure. However,  we
still present our theoretical results from $1.87$ to $5.5$~GeV to show
the interaction mechanism at higher energies.
\begin{figure}[h!]
\begin{center}
  \includegraphics[ bb=170 230 870 550,scale=0.9,clip]{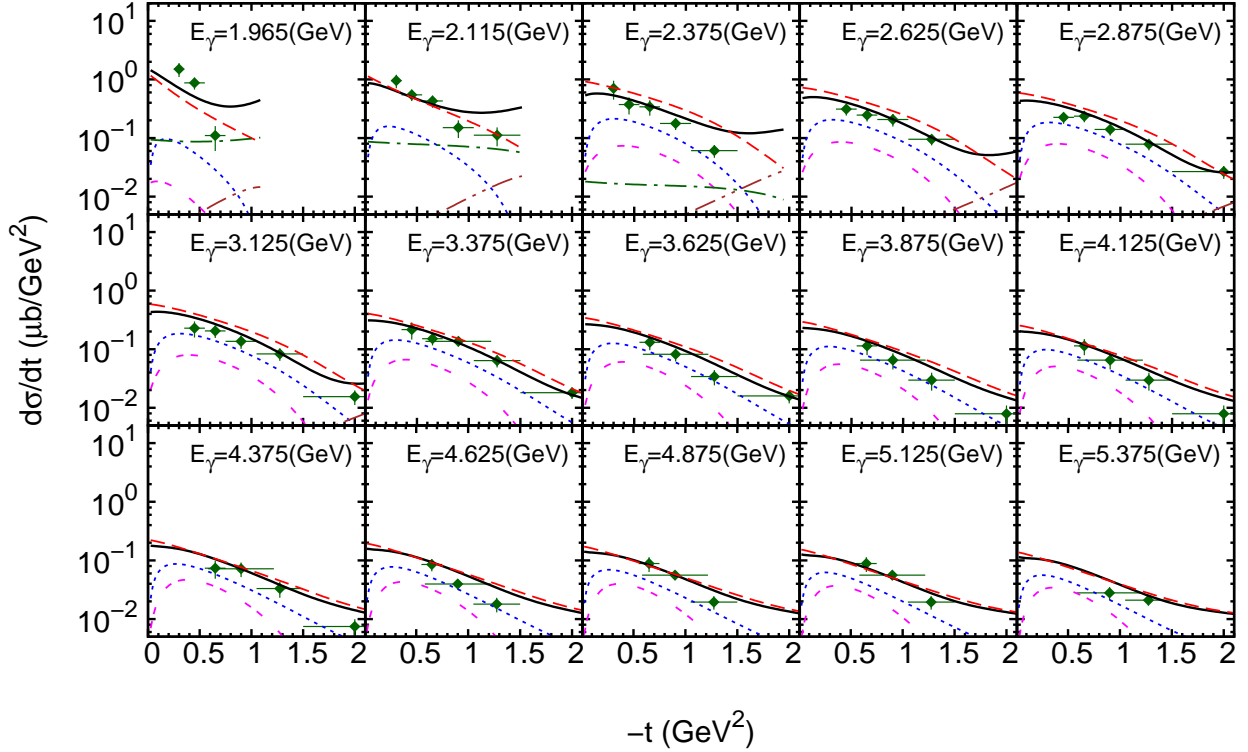}
\end{center}
  \caption{(Color online) The differential cross section $d\sigma/dt$ with the
	  variation of $-t$. The full diamond (darkgreen) and open
	  square (cyan) are for the preliminary $eg3$
	  data~\cite{Zhao2011}.
	  The notation as in Fig.~\ref{Fig: CLAS13}}\label{Fig: CLAS10dcsp}
\end{figure}
As shown  in Fig.~\ref{Fig: CLAS10dcsp}, our theoretical results are consistent with the preliminary
$eg3$ data especially the later one. The contribution from the
contact term plays most important role at the energies from threshold up to 5.5~GeV
while the contributions from nucleon resonances decreases rapidly at
the photon energy large than the energy point corresponding to the
Breit-Wigner mass.

\subsection{Total cross section}

\begin{figure}[h!]
\begin{center}
  \includegraphics[ bb=0 260 380 510 ,scale=1.,clip]{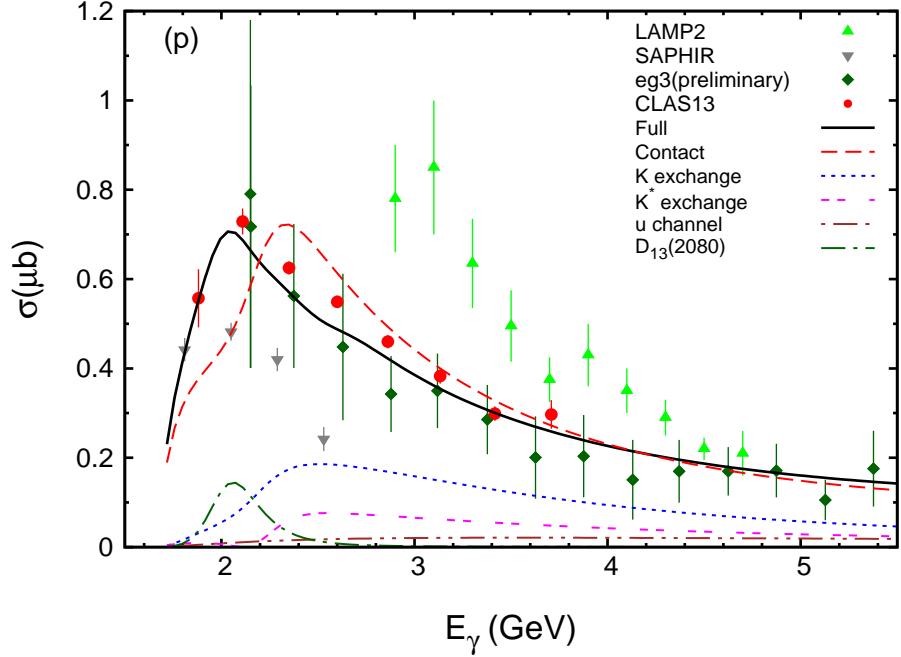}
\end{center}
  \caption{(Color online) Total cross section $\sigma$ with the
	  variation of the energy of photon $E_\gamma$.
 The notations for the theoretical results as in Fig.~\ref{Fig: CLAS13}.
  The data are from
  Refs.~\cite{Moriya:2013hwg,Zhao2011,Barber:1980zv,Wieland:2010cq}.}\label{Fig: TCS}
\end{figure}

The total cross section can be
obtained by integrating up the differential cross section. The
uncertainties will be introduced in the extrapolation of the
experimental differential cross section at measured angles to
the values at all angles. Hence, we do not include the experimental
total cross section in the fitting procedure. Our results calculated
with the determined parameters are shown in
Fig.~\ref{Fig: TCS} and compared with the experimental data.
Besides the CLAS13 data, the old data by the SAPHIR experiment at the low energy~\cite{Wieland:2010cq}
and the LAMP2 experiment at high energy~\cite{Barber:1980zv}  and  preliminary
$eg3$ data are also presented.
As shown in Fig.~\ref{Fig: TCS}, our results are close to the
experimental data.
The contact term is dominant at all energies considered in the current work.
The contribution from $N(2120)$ is important at energies about 2.1
GeV.

\section{Summary}

To determine the interaction mechanism of the photoproduction of
$\Lambda$(1520), the high precision data with large range of kaon
photoproduction angles are required.  In this work we investigated the
photoproduction of $\Lambda$(1520) in a Regge-plus-resonance
approach based on the new high precision CLAS13 data . The result
suggests that the contact terms are dominant in the photoproduction of
$\Lambda(1520)$.  The
Regge trajectory is found essential to reproduce the experimental data
at forward angles especially at high energies. To describe the
behavior of the differential cross section at backward angles, the
$\Lambda$ intermediate $u$ channel  contribution should be included.

The nucleon resonance $N(2120)$ with $3/2^-$ is essential to reproduce
the bump structure of near 2.1 GeV. In the new version of
PDG~\cite{PDG}, a state $N(1875)$ with $3/2^-$ is suggested. If the
third $[N\textstyle{3\over 2}^-]_3$ state is set to $N(1875)$, the
contribution from $[N\textstyle{3\over 2}^-]_3$ state with mass 1.875
GeV should be small considered the threshold of the $\Lambda$(1520)
photoproduction about 2.01 GeV.  Moreover, other nucleon resonances
predicted in the constituent quark model have small contributions as
shown in the current work.  So, the assignment of
$[N\textstyle{3\over 2}^-]_3$ as $N(1875)$ will lead to a small
$D_{13}$ partial wave contribution in the $\Lambda(1520)$ photoproduction.  However,
the results in this work and literatures [12,14] suggest that a
$D_{13}$ state about 2.1 GeV is essential to reproduce the
experimental data. Hence, the $[N\textstyle{3\over 2}^-]_3$ should be
assigned to $N(2120)$, not $N(1875)$.     

\section*{Acknowledgement}

This project is partially supported by the Major State
Basic Research Development Program in China (No. 2014CB845405),
the National Natural Science
Foundation of China (Grants No. 11275235, No. 11035006)
and the Chinese Academy of Sciences (the Knowledge Innovation
Project under Grant No. KJCX2-EW-N01).

\end{document}